\documentclass[11pt]{article}

\usepackage[margin=1in]{geometry}
\usepackage{amsmath}
\usepackage{amssymb}
\usepackage{graphicx}
\usepackage[colorlinks=true,linkcolor=blue,citecolor=blue,urlcolor=blue]{hyperref}

\title{A machine-learned dynamical phase diagram of the one-dimensional
nonlinear Schr\"odinger equation with quasiperiodic disorder and a static
field}

\author{
  Marcos P\'erez \\[2pt]
  \small Instituto de F\'isica, Universidade Federal do Rio Grande do Sul (UFRGS), Porto Alegre, Brazil \\
  \small \texttt{marcos.perez@ufrgs.br}
}
\date{}

\begin{document}
\maketitle

\begin{abstract}
We map the transport regimes of a wave packet evolving under the
one-dimensional discrete nonlinear Schr\"odinger (Gross--Pitaevskii) equation
with a quasiperiodic Aubry--Andr\'e potential and a static (Stark) field, as a
function of three controls: self-interaction $g$, field $F$, and quasiperiodic
strength $\lambda$, for two initial states (single-site delta and Gaussian).
Instead of one expensive long-time simulation per parameter point, we run
3600 short simulations ($t_{\max}=2.5\times10^{3}$, averaged over four
disorder-phase realizations), reduce each to eight physically motivated
dynamical features, and partition parameter space with Gaussian-mixture
clustering. Clustering finds the \emph{boundaries} between regimes without
supervision; each cluster's physical \emph{name} is then assigned by a
hand-calibrated threshold rule, and an ablation shows the clustering step
measurably smooths those boundaries rather than merely relabeling them. The
five resulting regimes (ballistic, localized, oscillatory-localized,
subdiffusive, and self-trapped) are validated against 100 long runs
($t_{\max}=10^{4}$) on a lattice eight times larger than the short sweep. The
lattice is sized specifically so the pre-boundary validation window is
genuinely longer than the short-run training window for every regime,
including the fastest-spreading one, which a naively matched lattice size
fails to guarantee. The short-time clusters predict the long-time asymptotic
spreading exponent ($\alpha_\infty = 2.01\pm0.07$ ballistic, $0.12\pm0.15$
localized, $0.23\pm0.17$ oscillatory-localized, $0.28\pm0.16$ subdiffusive,
consistent with the weak-chaos prediction $\alpha=1/3$) and, independently,
the long-time retention
$\Pi_0=0.72\pm0.32$ for the self-trapped cluster, whose $\alpha_\infty$ alone
is not diagnostic since it measures a radiated tail rather than the trapped
core. The initial-state contrast is striking and matches
Larcher--Dalfovo--Modugno: the delta state develops a broad self-trapping
wedge (onset at $g=3.907\pm0.017$ at $\lambda=0$, located by a dedicated
fine-resolution cut) that invades both the ballistic and localized regions
as $g$ grows, while the Gaussian state shows no self-trapping anywhere on
the same grid, instead opening a subdiffusive corridor along the
Aubry--Andr\'e critical line that widens with $g$. The scheme yields a full
$15\times15$ phase map for approximately 13 minutes per slice-and-state on a
laptop (8 cores), a regime in which direct asymptotic simulation
($t\gtrsim10^{6}$ per point in the literature) is prohibitive.
\end{abstract}


\section{Introduction}
\label{sec:intro}

A single one-dimensional lattice equation contains three of the most studied
mechanisms that arrest or promote the transport of a quantum wave. Quasiperiodic
("incommensurate") disorder, realized by the Aubry--Andr\'e--Harper potential,
produces in its linear limit a self-dual metal--insulator transition: all
single-particle eigenstates are extended below a critical potential strength
and exponentially localized above it \cite{aubry1980}. Self-interaction, the
mean-field Gross--Pitaevskii cubic nonlinearity, is known to destroy
single-particle (Anderson or Aubry--Andr\'e) localization and replace it with
slow subdiffusive spreading. At strong interaction it can instead self-trap a
wave packet \cite{pikovsky2008,kopidakis2008,flach2009}. A static (dc) field
produces, in the linear lattice, a Wannier--Stark ladder with spatial
localization and Bloch oscillations. Nonlinearity competes with this too,
damping and eventually destroying the oscillations and delocalizing the
packet \cite{krimer2009}.

Each mechanism on its own produces a distinct transport regime: ballistic,
localized, subdiffusive (weak- or strong-chaos), self-trapped, or
Bloch-oscillating. The combined three-parameter problem $(g,F,\lambda)$ has,
to our knowledge, not been mapped as a single dynamical phase diagram. A main
reason is cost: distinguishing a subdiffusion exponent $\alpha\approx1/3$
from $1/2$, or from true localization, requires integrating over many
decades in time \cite{flach2010,skokos2009}.

The central idea explored here is to replace one long, expensive simulation
per grid point with a large number of short, cheap simulations, extract a
compact vector of dynamical features from each, and use clustering to
partition parameter space into dynamical phases, validated against a much
smaller number of expensive long-time ``ground truth'' runs. This connects to
a growing literature that uses machine learning to detect phases of matter
without full prior knowledge of an order parameter
\cite{carrasquilla2017,vannieuwenburg2017,rodrigueznieva2019}; here the
``phases'' are dynamical (transport regimes of a driven, disordered,
nonlinear system) and the learner acts on time-series features of a single
wave function rather than on a static configuration.

We are careful, in what follows, to state precisely what this pipeline does
and does not discover without supervision. Clustering finds the
\emph{boundaries} between regimes in an unsupervised way; the physical
\emph{identity} of each region is still supplied by a small set of
hand-calibrated threshold rules on the same features, applied to each cluster
by majority vote. We quantify explicitly what the clustering step adds over
those rules alone (Sec.~\ref{sec:naming}), and we validate the resulting map
not only through the transport exponent but, for the one regime where that
exponent is not by itself diagnostic, through a second, independent long-time
observable (Sec.~\ref{sec:validation}).

The rest of the paper is organized as follows. Section~\ref{sec:model}
defines the model and the two initial states. Section~\ref{sec:method}
describes the integrator and validates it against known limits.
Section~\ref{sec:pipeline} sets out the full machine-learning pipeline:
sampling, features, clustering and its stability, what the naming step does
and does not learn, and the design of the long-time validation. Results are
in Section~\ref{sec:results}: the phase diagrams, a fine-resolution check of
the self-trapping onset, validation against long-time ground truth, the
feature-space geometry, representative dynamics, and cost. Section
\ref{sec:discussion} discusses what the maps show and where the method's
confidence is limited, Section~\ref{sec:outlook} lists directions for future
work, and Section~\ref{sec:conclusion} closes with a summary.

\section{Model}
\label{sec:model}

We integrate the discrete nonlinear Schr\"odinger / Gross--Pitaevskii chain
\begin{equation}
\label{eq:dnls}
i\,\dot\psi_n = -\left(\psi_{n+1}+\psi_{n-1}\right) + \varepsilon_n\,\psi_n +
g\,|\psi_n|^2\,\psi_n,
\qquad
\varepsilon_n = \lambda\cos\!\left(2\pi\beta n+\phi\right) + F\,n,
\end{equation}
with hopping normalized to unity, $\hbar=1$, $\beta=(\sqrt5-1)/2$ (the inverse
golden mean), and total norm $\sum_n|\psi_n|^2=1$. Here $\lambda$ is the
quasiperiodic (Aubry--Andr\'e) strength, $\phi$ a disorder-realization phase,
$F$ the static field, and $g$ the self-interaction ($g>0$: repulsive /
defocusing, the Bose--Einstein-condensate convention).

Equation~\eqref{eq:dnls} contains three well-studied limits that anchor our
validation (Sec.~\ref{sec:validation-solver}): $g=F=0$ is the linear
Aubry--Andr\'e model, with its metal--insulator transition at $\lambda=2$
\cite{aubry1980}; $F=0$ with $g,\lambda\neq0$ is the interacting quasiperiodic
chain of Larcher, Dalfovo and Modugno \cite{larcher2009,larcher2012};
$\lambda=0$ with $F,g\neq0$ is the nonlinear Stark ladder of Krimer,
Khomeriki and Flach \cite{krimer2009}.

We consider two initial states: a single-site \emph{delta} excitation,
$\psi_n(0)=\delta_{n,0}$, and a broad \emph{Gaussian},
$\psi_n(0)\propto\exp[-n^2/(2w^2)]$ with $w=4$, both normalized. The contrast
between them is physically diagnostic: a compact initial state is far more
prone to self-trapping than a broad one \cite{larcher2009}.

At logarithmically spaced times we record the center of mass
$\bar n=\sum_n n\rho_n$ (with $\rho_n=|\psi_n|^2$), the second moment
$m_2=\sum_n(n-\bar n)^2\rho_n$, the participation number
$P=1/\sum_n\rho_n^2$, the compactness index $\zeta=P^2/m_2$
\cite{skokos2009}, and the residual peak norm
$\Pi_0=\sum_{|n|\le2}\rho_n$. The spreading exponent
$\alpha=\mathrm d\log m_2/\mathrm d\log t$ is the primary classifier of
transport: $\alpha\approx0$ is localized, $\alpha\approx1/3$ weak-chaos and
$\alpha\approx1/2$ strong-chaos subdiffusion \cite{pikovsky2008,flach2009,
skokos2009,flach2010,laptyeva2010}, $\alpha\approx1$ diffusive, and
$\alpha\approx2$ ballistic. Interaction-induced subdiffusion in a
quasiperiodic lattice has been observed experimentally with a $^{39}$K
condensate \cite{lucioni2011}.

\section{Numerical method}
\label{sec:method}

Equation~\eqref{eq:dnls} is integrated with a second-order split-step Fourier
(Strang) scheme: the hopping term is diagonal in $k$-space (dispersion
$-2\cos k$) and applied by FFT; the on-site-plus-nonlinear term is diagonal
in real space and applied as an exact per-site phase rotation, exact because
$|\psi_n|$ is conserved during that substep. Each substep is unitary on the
lattice, so the total norm is conserved to machine precision
($\sim10^{-12}$ observed) independent of the time step $\Delta t$. The total
energy (including the periodic wrap bond, to match the FFT topology exactly)
drifts as $O(\Delta t^2)$, measured as $5.6\times10^{-3}\to2.0\times10^{-4}$
for $\Delta t=0.1\to0.025$, and this drift serves as the accuracy monitor.
The short
sweep and long-time validation both use $\Delta t=0.1$; the short sweep uses
lattice size $N=2048$ and the long-time validation $N=16384$
(Sec.~\ref{sec:validation-design}).

Because the FFT boundary is periodic, every trajectory is truncated, for
feature extraction, at the first time the summed density in the outermost 16
sites on each side exceeds $10^{-4}$ (``edge-aware'' windowing); the
center-of-mass excursion and running maximum of $m_2$ are additionally
tracked at every integration step, independent of the log-spaced snapshot
grid, so they are immune to aliasing of Bloch oscillations against sparse
sampling.

\label{sec:validation-solver}

\begin{figure}[t]
\begin{center}
\includegraphics[width=0.95\textwidth]{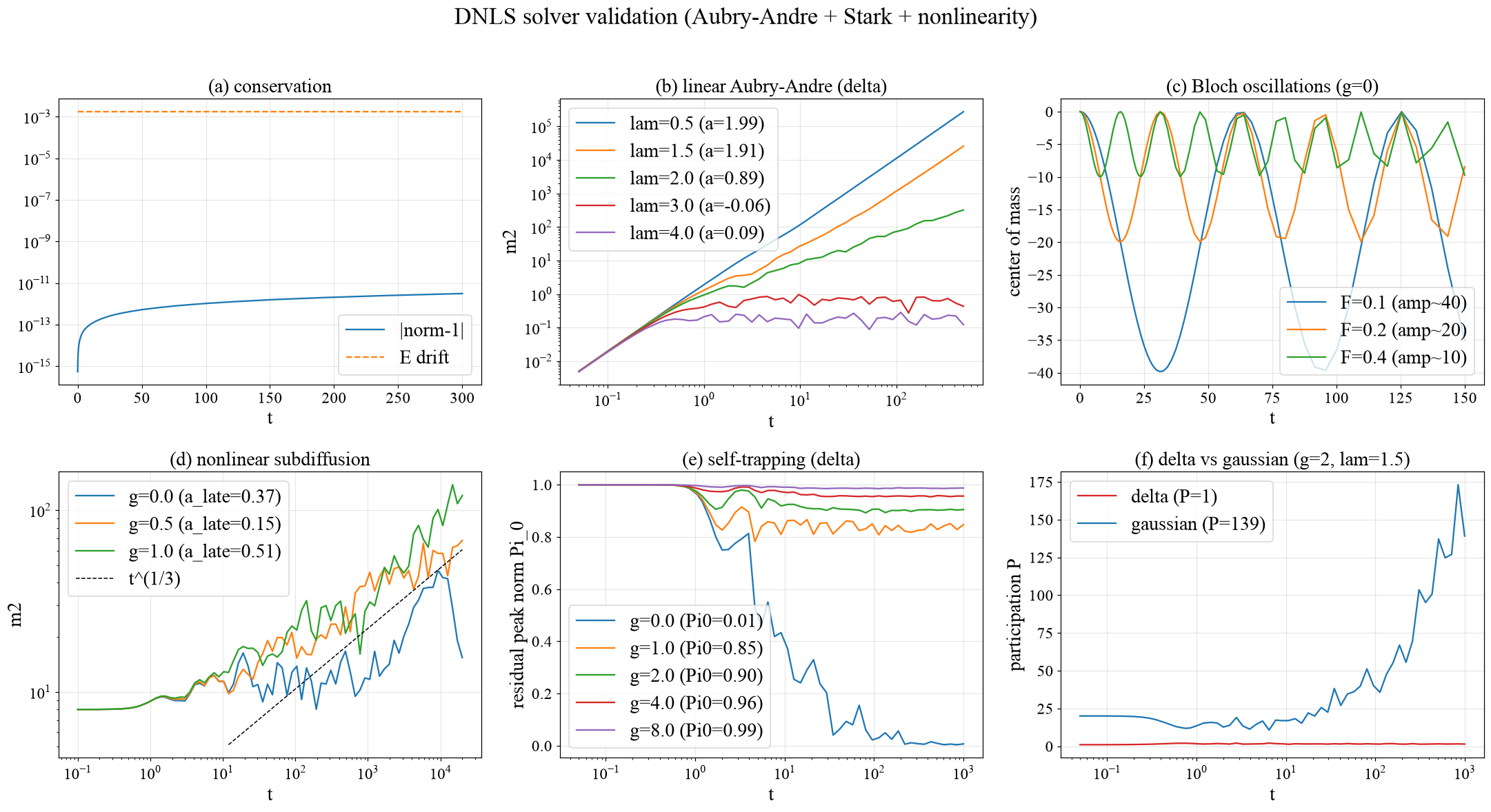}
\end{center}
\caption{Solver validation: (a) norm and energy conservation; (b) the linear
Aubry--Andr\'e metal--insulator transition at $\lambda=2$; (c) Bloch
oscillations with center-of-mass amplitude $4/F$; (d) interaction-induced
destruction of quasiperiodic localization; (e) self-trapping of a delta
state at strong interaction; (f) the delta-vs-Gaussian self-trapping
contrast of Larcher, Dalfovo and Modugno \cite{larcher2009}.}
\label{fig:solver-validation}
\end{figure}

Figure~\ref{fig:solver-validation} summarizes five checks against known
limits of Eq.~\eqref{eq:dnls}. (a) Conservation, as above. (b) The linear
Aubry--Andr\'e transition: $\alpha=1.99,\,1.91$ for $\lambda=0.5,\,1.5$
(extended), $\alpha\approx0$ with saturated $m_2$ for $\lambda=3,\,4$
(localized), with the crossover at $\lambda=2$ as expected. (c) Bloch
oscillations at $g=0$: the center-of-mass swing equals $4/F$ to three
significant digits (40.0, 20.0, 10.0 for $F=0.1,\,0.2,\,0.4$) with bounded
$m_2$. (d) Interaction destroys quasiperiodic localization: at $\lambda=2.5$,
$m_2(t=2\times10^4)$ grows from 15 ($g=0$) to 120 ($g=1$). (e) Self-trapping
of a delta state: $\Pi_0\to0.85$--$0.99$ for $g=1$--$8$ at $\lambda=1.5$,
while a broad Gaussian at the same parameters delocalizes instead
($P=1$ vs.\ $139$), matching the Larcher--Dalfovo--Modugno contrast
\cite{larcher2009}.

\section{Machine-learning phase-mapping pipeline}
\label{sec:pipeline}

\label{sec:sampling}

We sample two $15\times15$ parameter slices: slice A, $(g,\lambda)\in[0,4]^2$
at $F=0$ (interaction vs.\ quasiperiodic disorder), and slice B,
$(g,F)\in[0,4]\times[0,0.6]$ at $\lambda=1$ (interaction vs.\ static field).
Both slices are run for both initial states, with features averaged over
four quasiperiodic phase realizations $\phi\in\{0.7,\,2.27,\,3.84,\,5.41\}$
(spread evenly over $[0,2\pi)$). This gives $2\times2\times15\times15\times4
=3600$ short runs at $t_{\max}=2.5\times10^3$, costing approximately 13
minutes per slice-and-state (900 runs) on 8 laptop cores.

The phase-to-phase spread of the averaged features is tracked explicitly
(saved alongside the mean in each dataset file) rather than assumed small.
The median relative spread is 11--14\% for the Gaussian initial state but
25--36\% for the delta state. Single-site excitations are noticeably more
sensitive to the disorder phase than broad ones, as expected, since a compact
packet samples a much smaller stretch of the quasiperiodic potential. Four
realizations are not enough to remove this spread entirely; boundary
locations reported below for the delta panels should be read as accurate to
this precision.

\label{sec:features}

Each trajectory is reduced, on its pre-boundary window, to eight features
used for clustering: the late-window log-log slopes $\alpha_{m_2}$ and
$\alpha_P$ of $m_2$ and $P$; the final values $\log_{10}m_2$ and
$\log_{10}P$; the peak-drop $1-\Pi_0^{\rm fin}/\Pi_0^{\rm ini}$ (the
fraction of the initial central norm that has left the origin, an
initial-state-independent retention measure); a Bloch score, the per-step
center-of-mass excursion normalized by packet width; the final
$|\bar n|$; and the growth ratio $\log_{10}(m_2^{\rm fin}/m_2^{\rm early})$.
Two additional quantities, the compactness index and the raw final peak
norm, are exact linear combinations of these eight. They are excluded from
clustering, though retained for reporting, to keep the standardized feature
matrix full rank.

\label{sec:clustering}

For each initial state separately (absolute feature scales differ between
delta and Gaussian and must not be pooled across states), the two slices are
pooled, features standardized, and a Gaussian mixture model fit with the
number of components $k$ selected by the Bayesian information criterion over
$k=3\ldots8$. Both initial states select $k=6$ (BIC $=-2552$ delta,
$-6728$ gaussian; silhouette 0.375 delta, 0.370 gaussian). This is a modest
separation, consistent with the boundaries reported below being genuine
physical crossovers rather than sharp transitions. We additionally check the
stability of the cluster assignment at this $k$ with a 30-fold bootstrap:
resample the runs with replacement, refit, relabel the full data set with
the resampled model, match cluster identities to the reference fit by
maximum overlap (a linear-sum assignment on the in-bag contingency table),
and record the fraction of points that keep their original label. This gives
75.5\% (delta) and 84.2\% (gaussian) mean stability, a real and quantified
limit on precision rather than an implicit assumption of a uniquely correct
$k$.

\label{sec:naming}

Only the \emph{boundaries} between regimes are found without supervision, by
the Gaussian mixture model. Each cluster's physical \emph{name} is assigned
by a majority vote, over its members, of a per-point threshold rule on
(peak-drop, growth ratio, $\alpha_{m_2}$, Bloch score). For example,
\emph{self-trapped} is defined as retaining more than 50\% of the initial
central norm while $m_2$ grows by more than $0.8$ decades (a radiating
nonlinear tail); the full rule set is given in~\ref{app:rules}. This is a
genuine limitation on any claim that the pipeline ``discovers phases without
supervision'': the taxonomy is supplied by hand, and only the partition of
parameter space into instances of that taxonomy is learned.

To make explicit what the clustering step contributes beyond the hand-built
rule, we compare the GMM-plus-rule phase map against the same rule applied
pointwise, with no clustering step at all, using a simple neighbor-disagreement
``roughness'' score: the fraction of grid cells whose label differs from the
majority label of their orthogonal neighbors, where a lower value means the
map is spatially more coherent. Clustering reduces roughness in all four
slice/state combinations: 0.080 vs.\ 0.089 (A/delta), 0.053 vs.\ 0.062
(A/gaussian), 0.031 vs.\ 0.120 (B/delta), and 0.080 vs.\ 0.160 (B/gaussian),
for GMM-plus-rule vs.\ rule-only respectively. The effect is largest in the
field-driven slice B, where per-point thresholding is noisiest. We also
report, per slice and state, the \emph{crossover fraction}: the share of
cells where the individual-point rule disagrees with its cluster's majority
name, marked as white dots in Fig.~\ref{fig:phasediag}. This is 16.4\%
(A/delta), 10.7\% (A/gaussian), 22.7\% (B/delta), and 39.1\% (B/gaussian).
The last of these is high enough to flag explicitly: nearly two-fifths of
the field-driven Gaussian slice carries a disagreement marker, so that
slice's labels should be read as the coarser, less confident half of the
map.

\label{sec:validation-design}

We validate the short-time phase map against 100 independent long runs
($t_{\max}=10^4$ nominal, on a $5\times5$ subgrid of each $15\times15$
slice/state combination). These runs use a lattice \emph{eight times larger}
than the short sweep, $N=16384$ vs.\ $2048$. This is a deliberate design
choice, not the naive one of simply matching the short sweep's lattice. At
$N=2048$, a ballistic packet's group velocity (bounded by 2 in these units)
carries it into the edge-truncation window by $t\approx500$, which is
\emph{shorter} than the short run's own $t_{\max}=2500$: a long-time
validation run at that lattice size would, for exactly the fastest-spreading
phase, be strictly less informative than the training data it is meant to
validate. At $N=16384$ the pre-boundary window for the same ballistic point
extends to $t\approx4300$ in a single measured case, comfortably past the
short-run horizon. Rather than assume the nominal $t_{\max}$ is always
reached, we record the achieved pre-boundary duration $t_{\rm valid}$ for
every long run and report it explicitly alongside the transport exponent
(Sec.~\ref{sec:validation}), together with the long-time residual peak norm
$\Pi_0$, which we use as a direct, independent check on self-trapping.

\section{Results}
\label{sec:results}

\label{sec:phasediagrams}

\begin{figure}[p]
\begin{center}
\includegraphics[width=\textwidth]{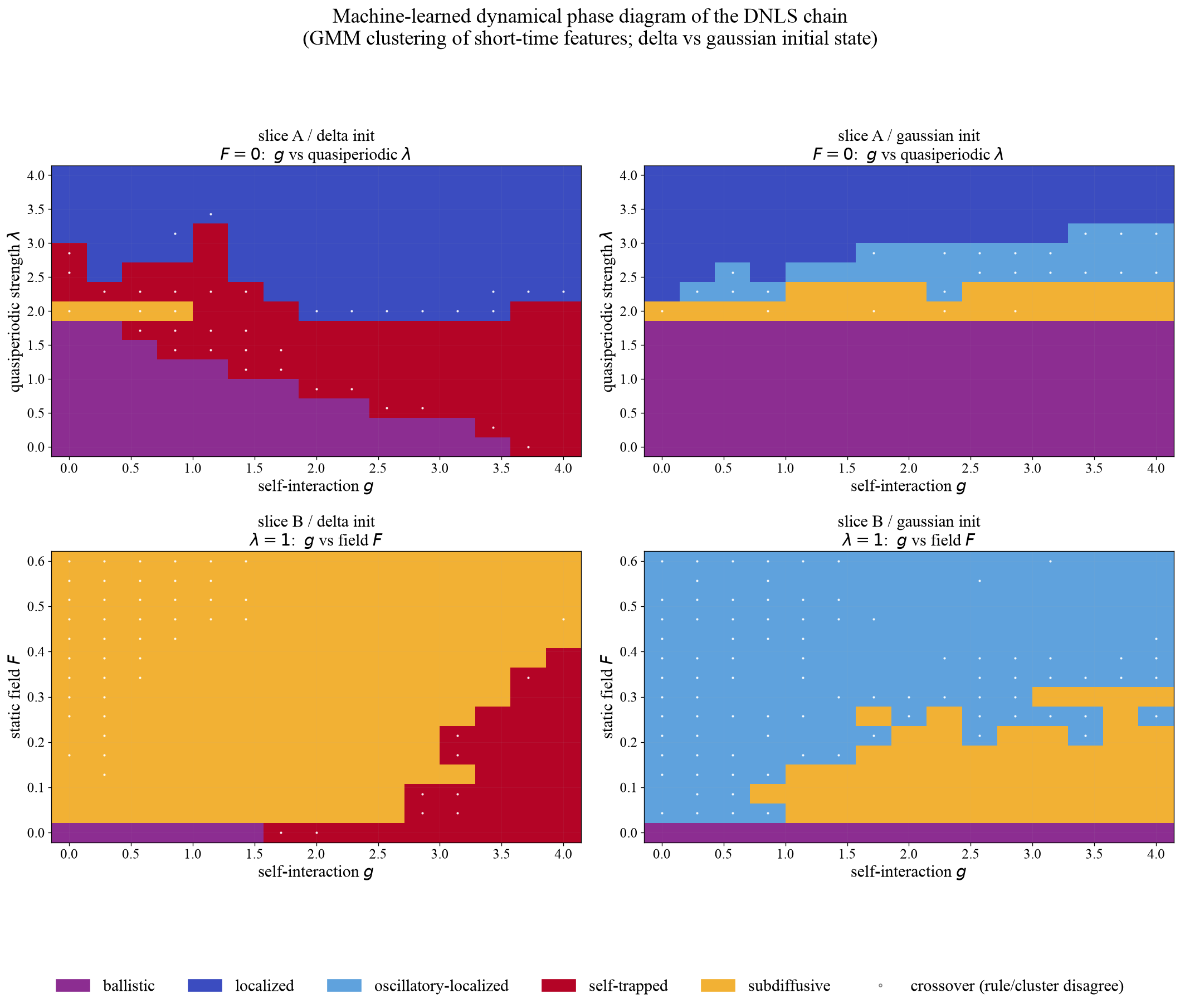}
\end{center}
\caption{Machine-learned dynamical phase diagrams. Rows: slices A
($F=0$, $g$ vs.\ $\lambda$) and B ($\lambda=1$, $g$ vs.\ $F$). Columns:
delta and Gaussian initial states. White dots mark crossover cells, where
the per-point threshold rule disagrees with its cluster's majority-vote
name (Sec.~\ref{sec:naming}).}
\label{fig:phasediag}
\end{figure}

Figure~\ref{fig:phasediag} shows the four resulting phase maps. The grid
spacing is $\Delta g=0.286$ in both slices, $\Delta\lambda=0.286$ (slice A),
and $\Delta F=0.043$ (slice B); boundary locations quoted below are accurate
only to this resolution unless a finer cut is cited explicitly.

Slice A, Gaussian (the ``clean'' interacting Aubry--Andr\'e map) shows
three main bands: ballistic below $\lambda\approx1.9$ (the AA metal,
essentially $g$-independent), a thin \emph{oscillatory-localized} band
($\lambda\approx2.1$--$3.3$, widening with $g$), and localized above. The
oscillatory-localized band is interaction-induced breathing of an
AA-localized state; it is resolved as distinct from strict insulation once
the $k=6$ fit is used. A \emph{subdiffusive corridor} pinned to the critical
line $\lambda\approx2$--$2.5$, widening with $g$, separates the ballistic
band from the oscillatory-localized one: interaction progressively destroys
quasiperiodic localization from the critical region outward, exactly the
physics of \cite{larcher2009,larcher2012}.

Slice A, delta shows the same metal/insulator skeleton, but a broad
\emph{self-trapping wedge} replaces much of it. A dedicated fine
one-dimensional cut at $\lambda=0$ (61 points, $\Delta g=0.033$; see
Sec.~\ref{sec:fine-cut}) locates the onset at $g=3.907\pm0.017$, consistent
with, and far more precise than, the coarse grid's $g\approx3.6$--$3.7$. The
threshold then falls with disorder (to $g\approx1$ at $\lambda\approx1.5$,
and toward $g\to0^+$ near the transition), since disorder assists
self-trapping by pre-localizing the packet. The wedge's outline against the
ballistic region reproduces the known delta-state phenomenology
\cite{flach2009,larcher2009}. At $g=0$, $\lambda>2$, the map inherits a
crossover-flagged column: a linear localized delta beats between hybridized
sites, mimicking the oscillation/growth signature that the rule set uses to
detect self-trapping (all such cells carry the disagreement marker; see
Sec.~\ref{sec:discussion}).

Slice B ($\lambda=1$, field on): for the Gaussian, an
\emph{oscillatory-localized} (Bloch) region at weak interaction
($g\lesssim1$--$1.5$, all $F$) gives way, with increasing $g$, to
subdiffusive spreading, since nonlinearity destroys Stark/Bloch localization
\cite{krimer2009}; the boundary moves to larger $g$ as $F$ grows (stronger
ladders resist longer), though this boundary carries the slice's highest
crossover density and should be read qualitatively. For the delta, most of
the slice is subdiffusive (field-assisted slow leakage), with self-trapping
capturing the strong-$g$ corner, its threshold rising roughly linearly with
$F$ (from $g\approx1.5$ at $F\to0$ to $g\approx4$ at $F\approx0.4$). The
$F=0$ row of both panels correctly reverts to the slice-A phenomenology at
$\lambda=1$ (ballistic $\to$ self-trapped at large $g$ for the delta).

\label{sec:fine-cut}

Because $\lambda=0$ makes the quasiperiodic term in Eq.~\eqref{eq:dnls}
vanish identically, the self-trapping onset along $F=0,\,\lambda=0$ for the
delta state does not depend on the disorder phase $\phi$, so a single dense
one-dimensional scan suffices. We run 61 points over $g\in[2.5,4.5]$
($\Delta g=0.033$, roughly $9\times$ finer than the main grid) and locate the
threshold crossing of peak-drop through the same 0.5 level used by the
per-point rule (Sec.~\ref{app:rules}) by linear interpolation, giving
$g=3.907\pm0.017$.

\label{sec:validation}

\begin{figure}[t]
\begin{center}
\includegraphics[width=\textwidth]{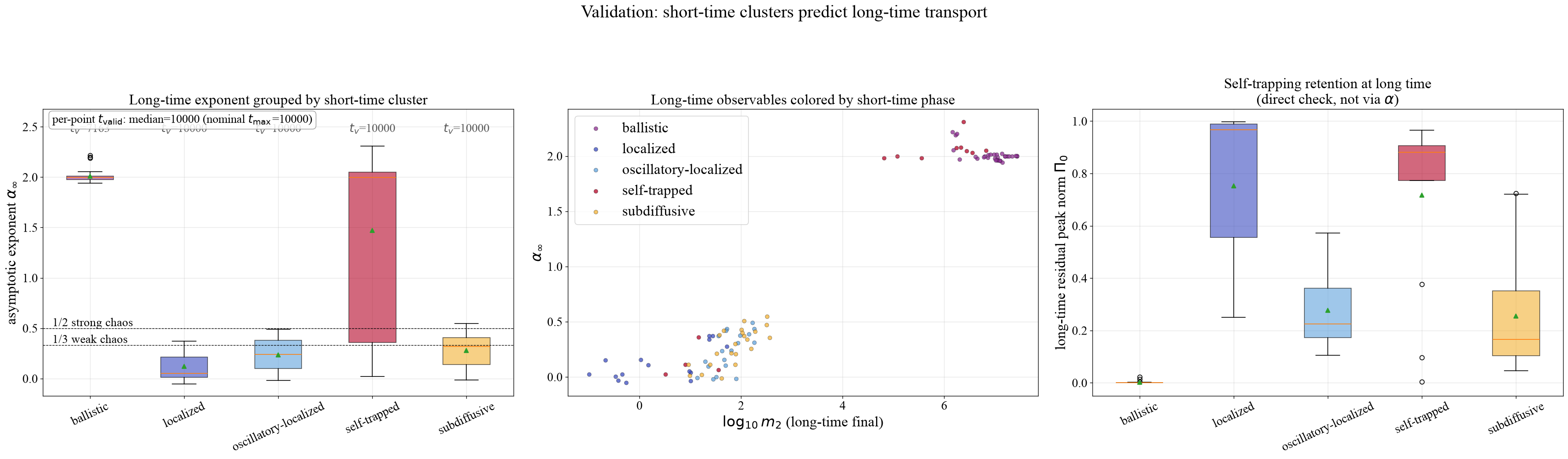}
\end{center}
\caption{Validation against long-time ground truth. Left: asymptotic
exponent $\alpha_\infty$ by short-time cluster (dashed lines mark the
weak- and strong-chaos values $1/3$ and $1/2$; $t_v$ annotations give the
median achieved $t_{\rm valid}$ per phase). Center: long-time
$(\log_{10}m_2,\alpha_\infty)$ colored by short-time phase. Right: long-time
residual peak norm $\Pi_0$ by short-time cluster, the direct self-trapping
check that does not rely on $\alpha_\infty$.}
\label{fig:validation}
\end{figure}

Table~\ref{tab:validation} summarizes the long-time behavior grouped by
short-time cluster; Fig.~\ref{fig:validation} shows the corresponding
distributions. As designed (Sec.~\ref{sec:validation-design}), the achieved
per-point $t_{\rm valid}$ reaches the full nominal $t_{\max}=10^4$ for every
phase except ballistic (median $7163$, still $\sim2.9\times$ the short-run
training horizon).

\begin{table}[t]
\caption{Long-time validation by short-time cluster. $n$: number of long runs
in this cluster (out of 100 total). $\alpha_\infty$: asymptotic spreading
exponent. $\Pi_0$: long-time residual peak norm. $t_{\rm valid}$: median
achieved pre-boundary duration (nominal $t_{\max}=10^4$).}
\label{tab:validation}
\begin{center}
\begin{tabular}{lcccc}
\hline
short-time phase & $n$ & $\alpha_\infty$ & $\Pi_0$ & median $t_{\rm valid}$\\
\hline
ballistic              & 31 & $2.012\pm0.067$ & $0.002\pm0.004$ & 7163\\
localized               & 15 & $0.120\pm0.147$ & $0.753\pm0.267$ & 10000\\
oscillatory-localized   & 19 & $0.234\pm0.170$ & $0.277\pm0.134$ & 10000\\
subdiffusive            & 22 & $0.281\pm0.163$ & $0.255\pm0.209$ & 10000\\
self-trapped            & 13 & $1.470\pm0.894$ & $0.717\pm0.320$ & 10000\\
\hline
\end{tabular}
\end{center}
\end{table}

Four of the five phases separate cleanly on $\alpha_\infty$ alone, and the
subdiffusive cluster's mean sits close to the weak-chaos value $1/3$
\cite{pikovsky2008,flach2009,skokos2009}. The self-trapped group's
$\alpha_\infty$ is, as expected, not diagnostic on its own
($1.47\pm0.89$, spanning most of the range from localized to ballistic):
$m_2$ of a self-trapped state measures the radiated tail, which is fast at
weak disorder and slow in the disordered background, while the trapped core
itself is invisible to $m_2$. This is no longer only an assertion: the
long-time residual peak norm $\Pi_0=0.717\pm0.320$, tracked independently of
$\alpha$, directly confirms that self-trapped points really do retain a
large fraction of their initial norm at $t=10^4$. This is second only to the
strictly localized cluster ($\Pi_0=0.753\pm0.267$) and well above every
delocalizing phase (subdiffusive $0.255$, oscillatory-localized $0.277$,
ballistic $0.002$). This is precisely why a multi-feature representation
(retention, growth, and exponent together), rather than the spreading
exponent alone, is needed to classify these dynamics. Both halves of that
claim, a bimodal $\alpha_\infty$ and a high $\Pi_0$, are now checked against
the same long-time data rather than one being inferred from the other.

\label{sec:featurespace}

\begin{figure}[t]
\begin{center}
\includegraphics[width=\textwidth]{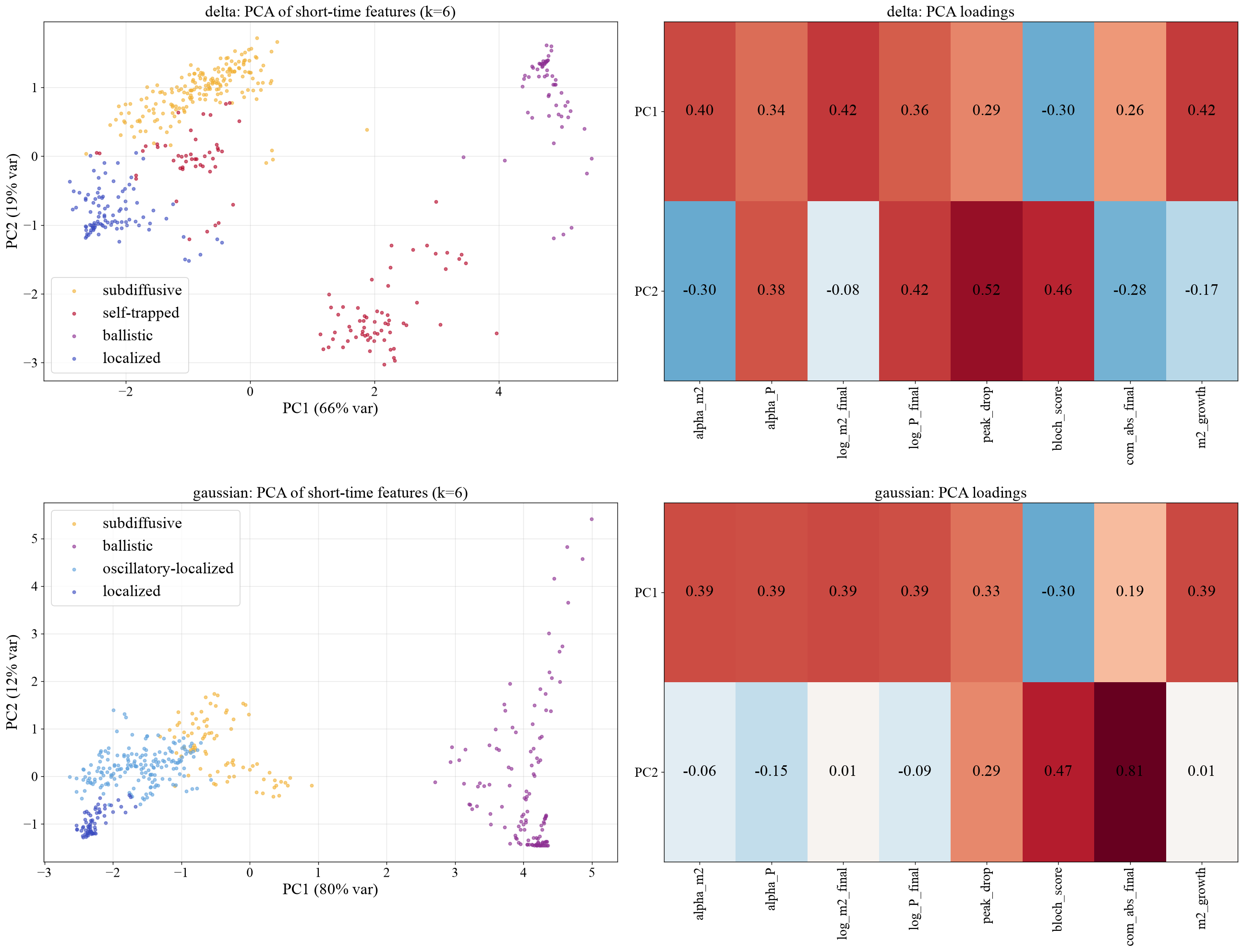}
\end{center}
\caption{PCA of the standardized short-time feature matrix, by initial
state (rows), with cluster identity colored (left) and PCA loadings
(right).}
\label{fig:featurespace}
\end{figure}

Figure~\ref{fig:featurespace} shows the first two principal components of
the standardized feature matrix, which together capture most of its variance
(66\%+19\% delta, 80\%+12\% gaussian). PC1 is dominated by the delocalization
block ($\alpha_{m_2}$, $\log m_2$, $\log P$, growth) and PC2 by the
retention/oscillation block (peak-drop, Bloch score, center-of-mass). The
two physically independent axes, ``how far/fast'' and ``what stayed
behind'', emerge from the clustering without being hand-specified, even
though the
cluster \emph{names} are (Sec.~\ref{sec:naming}). The self-trapped cluster in
the delta panel visibly splits into two PC1-separated clouds, the geometric
signature of the $\alpha_\infty$ bimodality quantified in
Sec.~\ref{sec:validation}.

\label{sec:portraits}

\begin{figure}[p]
\begin{center}
\includegraphics[width=\textwidth]{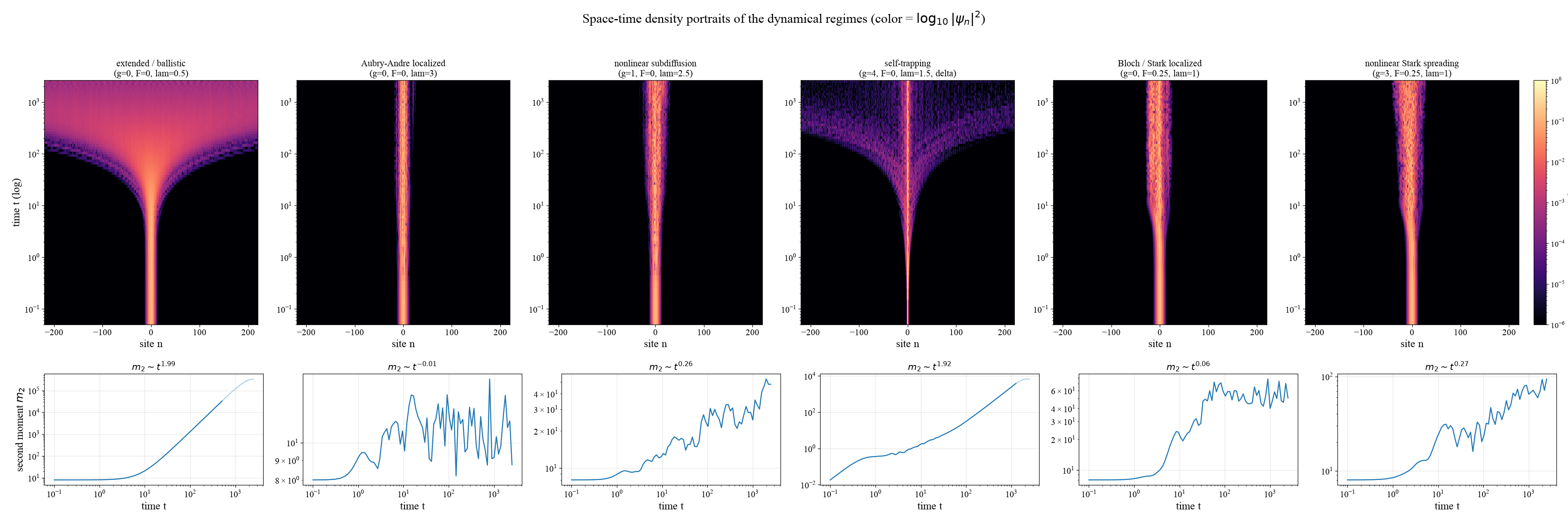}
\end{center}
\caption{Space-time density portraits $|\psi_n(t)|^2$ for one representative
parameter point per dynamical regime (top) and the corresponding $m_2(t)$
curve on log-log axes (bottom), with the fitted pre-boundary slope.}
\label{fig:portraits}
\end{figure}

Figure~\ref{fig:portraits} shows space-time density portraits
$|\psi_n(t)|^2$ for one representative parameter point per regime: a
ballistic light-cone ($m_2\sim t^{1.99}$); a frozen Aubry--Andr\'e insulator
($t^{-0.01}$); nonlinear subdiffusion ($t^{0.26}$ over the short window); a
self-trapped core with a radiated halo ($t^{1.92}$ for the tail, not the
core); a bounded Bloch/Stark packet ($t^{0.06}$); and nonlinear Stark creep
($t^{0.27}$).

\label{sec:cost}

The short-time map costs approximately 13 minutes per slice-state (225
points, 4 disorder phases, 900 simulations) on 8 laptop cores; using 2
phases instead, as in an earlier design, roughly halves this but undersamples
the disorder average (Sec.~\ref{sec:sampling}). The long-time validation
subset, on the $N=16384$ lattice required to make the validation meaningful
(Sec.~\ref{sec:validation-design}), costs approximately 14 minutes per
slice-state (25 points, 1 phase, 25 simulations). This is comparable
wall-clock time to the short sweep because it runs far fewer, far more
expensive simulations: each individual long run costs approximately
$39\times$ the
core-time of a short run, or approximately $10\times$ per grid point once the
short sweep's four-phase averaging is accounted for. Literature-grade
asymptotics ($t=10^6$--$10^8$ \cite{skokos2009,laptyeva2010}) remain
$10^2$--$10^4\times$ further out still. The clustering step itself is
seconds. Within the horizon actually achieved (Table~\ref{tab:validation}),
the short-time labels predicted the long-time regime; pushing the validation
horizon further specifically for the ballistic class (a still larger $N$) is
the main remaining cost knob.

\section{Discussion}
\label{sec:discussion}

\label{sec:disc-competition}
Quasiperiodicity localizes the chain. Interaction delocalizes the insulator
through subdiffusion, yet it arrests the metal through self-trapping for
compact initial states. The static field localizes the linear chain but
yields to interaction. The learned maps place all of these boundaries
quantitatively on the $(g,\lambda)$ and $(g,F)$ planes.

\label{sec:disc-initstate}
The choice between a delta and a Gaussian initial state is not a detail: it
changes the topology of the map, in particular the presence or absence of
the self-trapping wedge, consistent with \cite{larcher2009}. An experimental
protocol that quenches from a tight trap should therefore not be compared
directly with one that releases a broad condensate, since the two
correspond to different initial states in this model.

\label{sec:disc-limits}
Several limits are worth stating plainly. Cluster-level naming misassigns
heterogeneous clusters near boundaries; such cells are flagged rather than
hidden, and the flagged fraction is reported explicitly (10.7\%--39.1\%
depending on slice and state, Sec.~\ref{sec:naming}) rather than shown only
as scatter. Flagged cells concentrate along phase boundaries, in the
linear-beating column at $g=0,\,\lambda>2$, where they correctly signal
crossover physics (critical Aubry--Andr\'e anomalous transport at
$\lambda=2$), and throughout the field slice's
oscillatory-localized/subdiffusive boundary, which should be read as the
least confident region of the map.

The taxonomy itself is not learned; only the partition of parameter space
into that taxonomy is. Section~\ref{sec:naming} and its ablation state
clearly what the clustering step adds (smoother, more spatially coherent
boundaries than pointwise thresholding) and what it does not add (discovery
of the regime names).

The short window can also misread slowly developing subdiffusion as
localization near onset. The validation subset bounds this error, but only
out to $t_{\rm valid}\approx7\times10^3$ for the fastest, ballistic points.
Territory beyond that horizon remains genuinely unvalidated; this is now a
measured and reported limit rather than one hidden by an undersized
lattice.

In addition, $m_2$-based features alone cannot rank how much of a
self-trapped state is trapped, and are not diagnostic of self-trapping even
at long time ($\alpha_\infty=1.47\pm0.89$ for that cluster). Retention
features ($\Pi_0$) carry that information instead, and this is confirmed
directly against long-time data in Sec.~\ref{sec:validation} rather than
assumed.

Finally, bootstrap resampling shows 75.5\% (delta) and 84.2\% (gaussian)
cluster-label stability at the chosen $k$: a real but incomplete level of
confidence. The remaining $\sim$15--25\% of points are where the reported
boundaries should be expected to move under a different random seed or a
larger sample.

\label{sec:disc-metrics}
The question that motivated this study was which metrics to use, and the
answer follows community practice. The standard pair is $(m_2,P)$ with the
exponent $\alpha$ from $m_2\sim t^\alpha$ and the compactness index
$\zeta=P^2/m_2$ \cite{flach2009,skokos2009}. For self-trapping, the relevant
quantity is the residual norm $\Pi_0$, equivalently $1-$peak-drop
\cite{flach2009,larcher2009}; for Stark physics, it is the center-of-mass
excursion compared with $4/F$ \cite{krimer2009}. Our feature vector is
exactly this set, made edge-aware and independent of the initial state.

\section{Outlook}
\label{sec:outlook}

Several directions follow naturally. \emph{Active learning} could place new,
expensive samples where the Gaussian-mixture posterior is most uncertain
(the flagged cells) to refine boundaries at minimal additional cost.
\emph{Learned representations} could replace the engineered feature vector
with an encoder trained directly on the density time series, and the
discovered axes compared with the physically motivated $(\alpha,$
retention$)$ pair used here. A \emph{third axis}, dense 3D $(g,F,\lambda)$
sampling, is a direct extension, since the pipeline is already
dimension-agnostic. \emph{Beyond mean field}, TEBD/DMRG dynamics of the
corresponding Bose--Hubbard chain would test which of the boundaries
reported here survive quantum fluctuations, connecting to the many-body
localization literature under quasiperiodic potentials. Finally,
\emph{longer validation horizons} ($t\ge10^6$) on a small, targeted set of
boundary points would test the weak-to-strong-chaos crossover
\cite{laptyeva2010} against the short-time labels at a horizon this study
does not reach. This would be particularly useful along the ballistic
phase, where the present validation is limited to
$t_{\rm valid}\approx7\times10^3$ (Sec.~\ref{sec:cost}).

\section{Conclusion}
\label{sec:conclusion}

Short-time simulations, reduced to a small set of physically motivated
features and partitioned by Gaussian-mixture clustering, reproduce the gross
transport phenomenology of the quasiperiodic Stark--Gross--Pitaevskii chain
at a small fraction of the cost of direct long-time simulation, and the
resulting phase boundaries are corroborated on two independent long-time
observables, not just one, against expensive ground-truth runs sized
specifically to make that comparison meaningful. We have tried throughout to
state plainly where the method's confidence is high (the ballistic,
localized, and subdiffusive boundaries; the self-trapping retention check)
and where it is not (the field-slice boundaries, the ballistic validation
horizon, cluster stability at $k=6$), on the view that a phase diagram is
more useful to the reader with its uncertainty attached than without it.

\section*{Data availability statement}

The solver, feature-extraction, and clustering code
(\texttt{nlse\_phase/}), the analysis scripts that regenerate every figure
from scratch (\texttt{scripts/00--05}), and the resulting datasets
(short-sweep features with per-phase spread, long-time runs with per-point
$t_{\rm valid}$, and the fine-resolution boundary cut) are packaged for
reproduction and available at [repository URL].

\section*{Use of artificial intelligence}
An AI coding assistant was used during this work to help implement and
improve parts of the analysis code, and to help develop and clarify parts of
the discussion. All methods, results, and conclusions were reviewed by the
author, who takes full responsibility for the content of this manuscript.

\appendix
\section{Per-point threshold rule}
\label{app:rules}

Each grid cell's feature vector is assigned a rule label used for the
majority vote of Sec.~\ref{sec:naming}, based on the spreading exponent
$\alpha=\alpha_{m_2}$, the peak-drop $p$ (fraction of the initial central
norm that has left the origin), the Bloch score $b$, and the growth ratio
$\gamma=\log_{10}(m_2^{\rm fin}/m_2^{\rm early})$, applied in order:
\begin{itemize}
\item if $p<0.5$ and $\gamma>0.8$: \emph{self-trapped} (most of the initial
central norm is retained while a nonlinear tail radiates outward);
\item else if $p<0.5$ and $\gamma\le0.8$: \emph{localized} (central norm
retained, no real spreading);
\item else if $b>0.4$ and $\gamma<0.8$: \emph{oscillatory-localized} (a
bounded, oscillating packet: Bloch oscillation under a field, or
interaction-induced breathing of an Aubry--Andr\'e-localized state at
$F=0$);
\item else if $\alpha>1.4$: \emph{ballistic};
\item else if $\alpha>0.8$: \emph{diffusive};
\item else if $\gamma<0.4$ and $\alpha<0.5$: \emph{localized};
\item otherwise: \emph{subdiffusive} (slow but genuine delocalization).
\end{itemize}
Thresholds were calibrated against reference-regime runs (the solver
validation cases of Sec.~\ref{sec:validation-solver}) and are held fixed
across both parameter slices and both initial states.

\end{document}